\documentclass{Interspeech2024}
\usepackage{multirow}

\interspeechcameraready

\title{Learning Representation of Therapist Empathy in Counseling Conversation Using Siamese Hierarchical Attention Network}

\name{Dehua Tao$^1$, Tan Lee$^1$, Harold Chui$^2$, Sarah Luk$^2$}

\address{
  $^1$ Department of Electronic Engineering \quad 
   $^2$ Department of Educational Psychology\\The Chinese University of Hong Kong}
\email{dhtao@link.cuhk.edu.hk, tanlee@ee.cuhk.edu.hk, \{haroldchui, sarah\_luk\}@cuhk.edu.hk}


\begin{document}

\maketitle

\begin{abstract}
    
    Counseling is an activity of conversational speaking between a therapist and a client. Therapist empathy is an essential indicator of counseling quality and assessed subjectively by considering the entire conversation. This paper proposes to encode long counseling conversation using a hierarchical attention network. Conversations with extreme values of empathy rating are used to train a Siamese network based encoder with contrastive loss. Two-level attention mechanisms are applied to learn the importance weights of individual speaker turns and groups of turns in the conversation. Experimental results show that the use of contrastive loss is effective in encouraging the conversation encoder to learn discriminative embeddings that are related to therapist empathy. The distances between conversation embeddings positively correlate with the differences in the respective empathy scores. The learned conversation embeddings can be used to predict the subjective rating of therapist empathy.
    
\end{abstract}

\section{Introduction}
\label{sec:intro}

Counseling is a conversational activity involving a therapist and a client. The aim of counseling is to help clients with mental or emotional problems. In a counseling conversation, the therapist and the client take turn to speak. A speaker turn is defined as the time period during which only one person speaks. A typical counseling conversation lasts for a few tens of minutes and may contain hundreds of speaker turns. The empathy level expressed by the therapist during the conversation is an important indicator of counseling outcomes \cite{elliott2018therapist,moyers2013low,elliott2011empathy,miller2009toward}. It is defined as ``the therapist's sensitive ability and willingness to understand the client's thoughts, feelings, and struggles from the client's point of view" \cite{rogers1995way}. The acoustic properties of conversational speech between the therapist and the client have been shown to be related to the therapist empathy \cite{pasquale19_interspeech,xiao15b_interspeech,xiao14_interspeech,imel2014association,xiao13_interspeech}.


The sequence of speaker turns in a long conversation can be further divided into sub-sequences that contain multiple turns. These sub-sequences are referred to as sections in this paper. Each conversation section is expected to cover certain local speech characteristics in dyadic interactions \cite{jakkam2016, de2014investigating, levitan11_interspeech}. In this way, a counseling conversation can be represented by a hierarchical structure, i.e., consecutive turns form a section, and multiple sections constitute the conversation. Following the recent study \cite{tao22b_interspeech}, a hierarchical recurrent neural network is adopted to model counseling conversation at the turn and section levels. Two levels of attention mechanisms \cite{luong2015effective, bahdanau2014neural} are applied to capture importance weights of individual turns and sections. The hierarchical attention network carries the function of an encoder module to encode a variable-length conversation into a latent representation that can be linked to the quality of counseling and the therapist empathy.


In psychological studies, subjective assessment on empathy and/or counseling quality is usually done regarding the whole counseling conversation. That means only an overall rating of empathy is available but no labels of localized empathic events for a counseling conversation. We assume that conversations with polarized ratings of empathy contain more frequent and prominent empathic or non-empathic behaviors than those with indifferent values of empathy ratings. In this study, we select the conversations with empathy ratings on the two extremes and divide them into the low-empathy and high-empathy groups. We propose to use a Siamese network with contrastive loss \cite{norouzi2012hamming, hadsell2006dimensionality,chopra2005learning} to train an encoder to learn discriminative embeddings from the selected conversations. The input to the Siamese network comprises a pair of conversations. The contrastive loss is applied such that the distance between the two conversation embeddings would be minimized if they are from the same empathy group, e.g., both are high-empathy conversations, and maximized if they are from different groups.

Experimental results show that selecting counseling conversations with extreme empathy ratings is effective for training the conversation encoder with contrastive loss. The embeddings generated from the conversation encoder contain useful information about the therapist empathy. The distances between conversations in the embedding space are positively correlated with the differences in their empathy scores. Furthermore, the conversation embedding can be used to predict the therapist empathy in counseling.

Section \ref{sec:dataset} describes the counseling speech dataset used in this study. Section \ref{sec:system} explains the proposed Siamese conversation encoder with contrastive loss in detail. Section \ref{sec:expset} details the design of training data and test data, model configuration, and evaluation methods. Section \ref{sec:resdisc} presents and discusses the experimental results, followed by conclusion in Section \ref{sec:conc}.

\section{Counseling Speech Dataset}
\label{sec:dataset}

\begin{table*}[htb]
\centering
\caption{Summary of counseling conversations used in this study}
\label{tab:data}
\resizebox{1.0\linewidth}{!}{
\begin{tabular}{|ccc|ccc|ccc|}
\hline
\multicolumn{3}{|c|}{Speech time  per conversation (min)}                            & \multicolumn{3}{c|}{No. of tuns  per conversation}                               & \multicolumn{3}{c|}{Duration  of turns (sec)}                                                   \\ \hline
\multicolumn{1}{|c|}{Median} & \multicolumn{1}{c|}{Mean\textpm SD}    & Range of Min to Max & \multicolumn{1}{c|}{Median} & \multicolumn{1}{c|}{Mean\textpm SD} & Range of Min to Max & \multicolumn{1}{c|}{Median} & \multicolumn{1}{c|}{Mean\textpm SD}    & Range of 2nd to 98th percentile \\ \hline
\multicolumn{1}{|c|}{49.83}  & \multicolumn{1}{c|}{48.39\textpm 9.45} & 15.49 - 81.39       & \multicolumn{1}{c|}{283}    & \multicolumn{1}{c|}{302\textpm 137} & 54 - 781            & \multicolumn{1}{c|}{3.61}   & \multicolumn{1}{c|}{9.60\textpm 19.68} & 0.30 - 63.28                    \\ \hline
\end{tabular}}
\end{table*}

A counseling speech dataset named CUEMPATHY \cite{tao2022cuempathy} is used in this study. The dataset consists of $156$ audio recordings of counseling conversations involving $39$ therapists ($31$ female, $8$ male) and $39$ clients ($30$ female, $9$ male). Each of the $39$ therapists was paired with one client, thus forming $39$ unique therapist-client dyads. Each therapist-client dyad had $4$ conversations. The conversations were recorded during counseling practicums for therapist trainees at a university. The clients were adults who came to seek psychological assistance concerning relationship, emotion, stress, and self-esteem. All therapists and clients spoke Hong Kong Cantonese. The study was approved by the institutional review board, and informed consent was obtained from both the clients and therapists. Each counseling conversation is about $50$ minutes long. The audio recording of a conversation is divided into speaker-turn-based audio segments with an automatic speech-text alignment system \cite{tao2022cuempathy}. Table \ref{tab:data} gives a summary of the $156$ counseling conversations.

\begin{figure}[htb]
  \centering
  \includegraphics[width=0.8\linewidth]{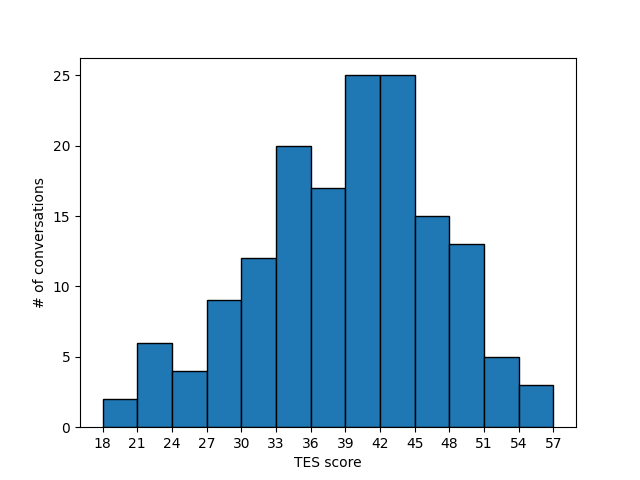}
  \caption{TES scores of 156 counseling conversations.}
  \label{fig:TES}
\end{figure}

The therapist's empathy level in each conversation was rated by a trained observer according to the Therapist Empathy Scale (TES) \cite{decker2014development}. The TES includes nine items that assess affective, cognitive, attitudinal and attunement aspects of therapist empathy. Each item is rated on a seven-point scale from $1=\textit{not at all}$ to $7=\textit{extremely}$. An example item is ``\textit{Expressiveness: A therapist’s voice demonstrates expressiveness when the therapist speaks with energy and varies the pitch of his or her voice to accommodate the mood or disposition of the client.}" The total TES score of a counseling conversation ranges from 9 to 63. A higher score indicates a higher level of therapist empathy. Eight observer raters with at least master's level training in counseling were recruited to rate the counseling conversations in this study. As a reliability check, about $40\%$ ($62$ conversations) of the recorded conversations were rated by two raters. The intraclass coefficient was $0.90$, indicating excellent interrater reliability \cite{cicchetti1994guidelines}. Figure \ref{fig:TES} shows the distribution of TES scores of the $156$ conversations. The scores range from $18$ to $56.5$ with the mean $38.80$\textpm$7.87$.

\section{Proposed System}
\label{sec:system}

\subsection{Conversation Encoder}
\label{ssec:encoder}

\begin{figure}[htb]
  \centering
  \includegraphics[width=0.75\linewidth]{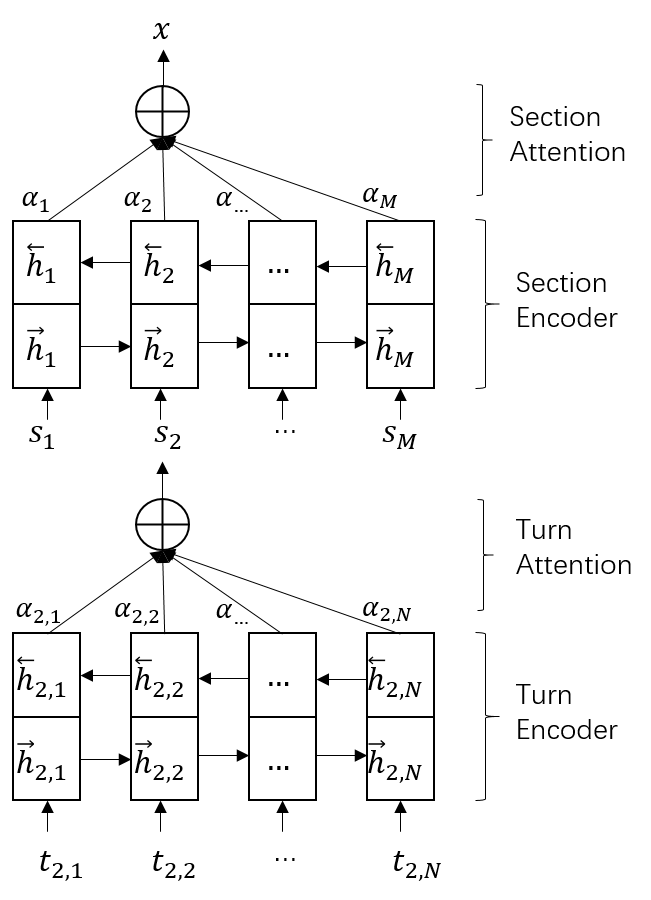}
  \caption{Illustration of the conversation encoder.}
  \label{fig:conenc}
\end{figure}

Similar to the hierarchical model in \cite{tao22b_interspeech}, two sub-structures at the turn and section levels are adopted as shown in Figure~\ref{fig:conenc}. For a counseling conversation, $N$ consecutive speaker turns from the therapist and client, denoted as $t_{i,j}$, $j=1,2, ..., N$, constitute a section. $N$ is a hyper-parameter to be determined by experiments. The conversation is divided into a series of sections denoted as $s_i$, where $i=1,2, ..., M$. 

\vspace{0.5em}
\noindent
\textbf{Encoder layer:} The function of the encoder layer (either turn level or section level) is to aggregate and transform a sequence of features into a hidden representation of the respective turn or section. The turn encoder is expressed by Eq. (\ref{eq:1}). A bidirectional gated recurrent unit (BiGRU) \cite{bahdanau2014neural, cho2014properties} encodes the turn sequence in each section. The hidden representation $h_{i,j}$ for turn $t_{i,j}$ is obtained by concatenating the forward hidden state $\overrightarrow{h_{i,j}}$ which reads the section $s_i$ from $t_{i,1}$ to $t_{i,N}$, and the backward hidden state $\overleftarrow{h_{i,j}}$ which reads from $t_{i,N}$ to $t_{i,1}$. The section encoder has a similar structure to the turn encoder. The hidden representation $h_i$ for section $s_i$ is obtained with a BiGRU in the same way, where $s_i$ and $h_i$ are analogous to $t_{i,j}$ and $h_{i,j}$, respectively.

\begin{equation}
\begin{aligned}
  &\overrightarrow{h_{i,j}} = \overrightarrow{\text{GRU}}(t_{i,j}, h_{i,{j-1}}) \\
  &\overleftarrow{h_{i,j}} = \overleftarrow{\text{GRU}}(t_{i,j}, h_{i,{j+1}}) \\
  &h_{i,j} = [\overrightarrow{h_{i,j}}, \overleftarrow{h_{i,j}}]
  \label{eq:1}
\end{aligned}
\end{equation}


\vspace{0.5em}
\noindent
\textbf{Attention layer:} To obtain a meaningful high-level embedding, attention mechanism is applied to put strong weights on important hidden state vectors in the input sequence. With turn-level attention, selected turns in a specific section are emphasized in contributing to the section embedding. Similarly, section representations that are significant to the conversation embedding are stressed through the section-level attention. The attention mechanism at turn level is expressed by Eq. (\ref{eq:2}). $u_{i,j}$ denotes the hidden representation of $h_{i,j}$ obtained by using a fully connected layer with \textit{tanh} activation function. The importance weight $\alpha_{i,j}$ of turn $t_{i,j}$ is obtained by calculating the similarity of $u_{i,j}$ and a turn-level context vector $u_t$, and normalizing it with a softmax function. The section embedding $s_i$ is computed as a weighted sum of turn-level hidden states. The same process is applied in the section-level attention. A section-level context vector denoted as $u_s$ is used to measure the importance of individual sections. Lastly, the conversation embedding $x$ is obtained by summarizing the information of all speaker turns.

\begin{equation}
\begin{aligned}
  &u_{i,j} = \text{tanh}(W_th_{i,j} + b_t) \\
  &\alpha_{i,j} = \frac{\text{exp}(u^\top_{i,j}u_t)}{\sum_j \text{exp}(u^\top_{i,j}u_t)} \\
  &s_i = \sum_j \alpha_{i,j}h_{i,j}
  \label{eq:2}
\end{aligned}
\end{equation}


\subsection{Siamese Network with Contrastive Loss}
\label{ssec:contraloss}

\begin{figure}[htb]
  \centering
  \includegraphics[width=0.75\linewidth]{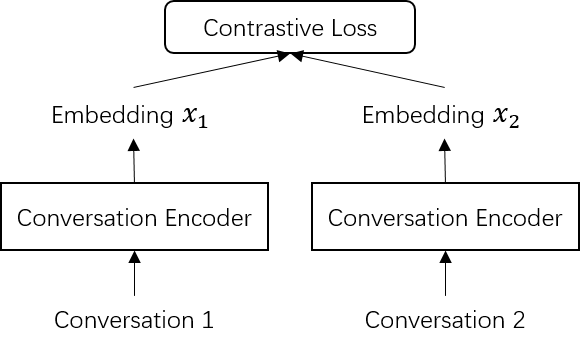}
  \caption{Siamese conversation encoder.}
  \label{fig:siamese}
\end{figure}

As shown in Figure \ref{fig:siamese}, the input to the Siamese conversation encoder is a pair of counseling conversations. It is expected that the empathic or non-empathic behaviors are similar in conversations with similar empathy scores while dissimilar in conversations with highly deviating scores. Let $x_1$ and $x_2$ denote the conversation embeddings encoded from the two input conversations. Let $y$ denote the binary label assigned to the input pair. $y=1$ if $x_1$ and $x_2$ have similar empathy scores, and $y=0$ if their scores deviate significantly. The contrastive loss is defined as

\begin{equation}
    L = \frac{1}{2}yd^2 + \frac{1}{2}(1-y)\{\text{max}(0, m-d)\}^2
    \label{eq:3}
\end{equation}

\noindent
where $d$ denotes the Euclidean distance between $x_1$ and $x_2$, and $m>0$ is a margin. $d$ should decrease if $y=1$ and increase if $y=0$.

\section{Experimental Setup}
\label{sec:expset}

\subsection{Training Data and Test Data}
\label{ssec:trndata}

Counseling conversations with empathy scores on the two extremes are selected and divided into the low-empathy and high-empathy groups. Two types of conversation pairs are created, including positive pairs with similar empathy scores from the same group and negative pairs with highly deviating scores from different groups. With such a design of training data, the Siamese conversation encoder is expected to learn discriminative embeddings that are related to the therapist empathy. Suppose that each of the two groups contains $K$ conversations. A total of $(2K^2-K)$ pairs of conversations can be formed, including $(K^2-K)$ positive pairs and $K^2$ negative pairs. Different values of $K = 10,15,20,25,30$ will be explored in our experiments.

Let $[C_1,...,C_{156}]$ denote the $156$ counseling conversations in the CUEMPATHY, arranged from the lowest to the highest in terms of the TES score. For a specific value of $K$, the conversations $[C_1,...,C_K]$ and $[C_{156-K+1},...,C_{156}]$ are used as training data. In addition, the $96$ conversations in the middle, namely $[C_{31}, ..., C_{126}]$, are used as test data. Their TES scores are from $32$ to $46$. The test set without any ``extreme" values allows for a reliable assessment of whether the conversation encoder is actually learning empathy-related information.

\subsection{Model Configuration}
\label{ssec:modelconf}

Each speaker turn is represented by the 88-dimensional eGeMAPS feature vector \cite{eyben2015geneva}. The turn-level features are normalized with respect to the mean and variance computed from all turns from the same speaker. The section size $N$ (number of turns) is a hyper-parameter to be determined by experiments. The number of sections ($M$) is then determined such that $N\times M$ is larger than the maximum number (i.e., $781$ as shown in Table \ref{tab:data}) of turns among all conversations. For example, $M$ is set to $200$ when $N=4$. The conversations without sufficient turns are zero-padded to the required number of turns.

In the conversation encoder, the number of GRU units are set to $64$ and $16$ at the turn and section level respectively. The turn and section context vectors have the dimensions of $128$ and $32$ respectively. The batch size is set to $64$. 
An Adam optimizer with learning rate of $0.001$ is used. The Siamese conversation encoder is trained for $30$ epochs.

\subsection{Evaluation Methods}
\label{ssec:eval}

The effectiveness of the conversation encoder is evaluated on the learned embeddings. Given two conversations that have very different TES scores, the distance between the embeddings computed from them is expected to be large. If their TES scores are close, their embeddings should have small distance. Ten conversations with the lowest TES scores (i.e., $[C_1,...,C_{10}]$) are used as the low-score references. Given an unseen test conversation, the Euclidean distance between its embedding and each of the reference embeddings is calculated. The average distance over the $10$ reference embeddings is computed. For all test conversations, the Pearson's correlation $\rho$ between the average distance and the respective TES scores is computed. The same procedure is repeated with $10$ reference conversations of the highest TES scores (i.e., $[C_{147},...,C_{156}]$).

To further validate the efficacy of the conversation encoder, experiments are carried out to use the learned embeddings to predict TES scores. The $96$ test conversations involve $35$ different therapist-client dyads. Leave-one-dyad-out cross-validation is carried out. The regression is implemented with the Support Vector Regression in scikit-learn \cite{pedregosa2011scikit}. The coefficient of determination $R^2$ is computed from the $96$ predictions.

To assess the effectiveness of the contrastive loss in guiding the conversation encoder to acquire discriminative embeddings associated with therapist empathy, we employ cross-entropy loss to train the encoder in a binary classification task that determines the level of therapist empathy during counseling conversation. The $R^2$ is computed for the test set predictions to establish a baseline for comparison.

\section{Results and Discussion}
\label{sec:resdisc}

\begin{table}[htb]
\centering
\caption{Correlation ($\text{p-value}<0.05$) and regression results for different $K$. ``Score Range" refers to the range of TES scores for the low-empathy and high-empathy groups.}
\label{tab:res}
\resizebox{0.95\linewidth}{!}{
\begin{tabular}{|c|c|cc|cc|}
\hline
\multirow{2}{*}{K} & \multirow{2}{*}{\begin{tabular}[c]{@{}c@{}}Score Range\\(Low vs. High)\end{tabular}} & \multicolumn{2}{c|}{$\rho$}                                                                                                                              & \multicolumn{2}{c|}{$R^2$$\uparrow$}                                                                      \\ \cline{3-6} 
                   &                                                                                       & \multicolumn{1}{c}{\begin{tabular}[c]{@{}c@{}}Low-score\\Reference\end{tabular}} & \begin{tabular}[c]{@{}c@{}}High-score\\Reference\end{tabular} & \multicolumn{1}{c}{Contrastive} & \begin{tabular}[c]{@{}c@{}}Cross-\\entropy\end{tabular} \\ \hline
10                 & \begin{tabular}[c]{@{}c@{}}(18 - 25.5) vs.\\ (49 - 56.5)\end{tabular}                 & \multicolumn{1}{c}{0.293}                                                         & -0.289                                                         & \multicolumn{1}{c}{0.020}       & 0.012                                                    \\ \hline
15                 & \begin{tabular}[c]{@{}c@{}}(18 - 28) vs.\\ (48.5 - 56.5)\end{tabular}                 & \multicolumn{1}{c}{0.341}                                                         & -0.345                                                         & \multicolumn{1}{c}{0.148}       & -0.009                                                   \\ \hline
20                 & \begin{tabular}[c]{@{}c@{}}(18 - 29) vs.\\ (48 - 56.5)\end{tabular}                   & \multicolumn{1}{c}{0.301}                                                         & -0.347                                                         & \multicolumn{1}{c}{\textbf{0.169}}       & -0.036                                                   \\ \hline
25                 & \begin{tabular}[c]{@{}c@{}}(18 - 31) vs.\\ (46.5 - 56.5)\end{tabular}                 & \multicolumn{1}{c}{0.323}                                                         & -0.390                                                         & \multicolumn{1}{c}{0.116}       & 0.059                                                    \\ \hline
30                 & \begin{tabular}[c]{@{}c@{}}(18 - 32) vs.\\ (46 - 56.5)\end{tabular}                   & \multicolumn{1}{c}{0.303}                                                         & -0.338                                                         & \multicolumn{1}{c}{0.130}       & 0.027                                                    \\ \hline\hline
20 (5)             & \begin{tabular}[c]{@{}c@{}}(22 - 31) vs.\\ (46.5 - 51)\end{tabular}                   & \multicolumn{1}{c}{0.253}                                                         & -0.334                                                         & \multicolumn{1}{c}{0.085}       & -0.023                                                   \\ \hline
20 (10)            & \begin{tabular}[c]{@{}c@{}}(26 - 32) vs.\\ (46 - 49)\end{tabular}                     & \multicolumn{1}{c}{\begin{tabular}[c]{@{}c@{}}Not \\ Significant\end{tabular}}    & -0.226                                                         & \multicolumn{1}{c}{0.011}       & -0.032                                                   \\ \hline
\end{tabular}
}
\vspace{-1.0em}
\end{table}

\subsection{Correlation between distances from references and TES scores}
\label{ssec:corr}

Table \ref{tab:res} shows the results of correlation coefficients between embedding distance measures and TES scores for different values of $K$ where we set $N=4$ and $M=200$. The value of margin $m$ in the contrastive loss is determined to be $2.0$ based on the experimental performance. Given a test conversation, its distances from the low-score and high-score references in the embedding space are both positively correlated with the difference between its TES score and the references' scores. This shows that the conversation encoder can encode conversations with similar scores into nearby embeddings and conversations with different scores into distant embeddings.

\vspace{-1.0em}
\begin{figure}[htb]
  \centering
  \includegraphics[width=0.8\linewidth]{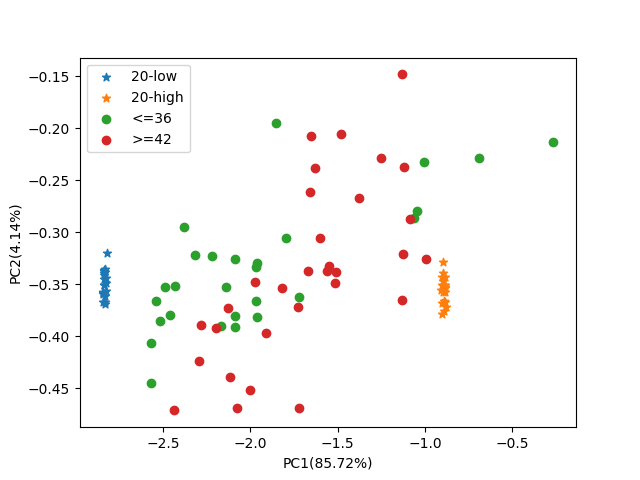}
  \caption{Visualization of conversation embeddings for K=20.}
  \label{fig:sessemd}
\end{figure}

For example, in the case of $K=20$, the first two principal components of conversation embeddings for the training data and part of the test data are visualized as in Figure \ref{fig:sessemd}. The blue and orange stars represent the conversations from the low-empathy and high-empathy training groups, respectively. The green and red dots represent the selected test conversations. Most of the green dots for conversations with TES scores of $36$ or lower are closer to the blue stars for low-empathy conversations. Similarly, most of the red dots for conversations with scores of $42$ or higher are near the orange stars for high-empathy conversations.

\subsection{Regression on conversation embeddings}

\begin{figure}[htb]
  \centering
  \includegraphics[width=0.8\linewidth]{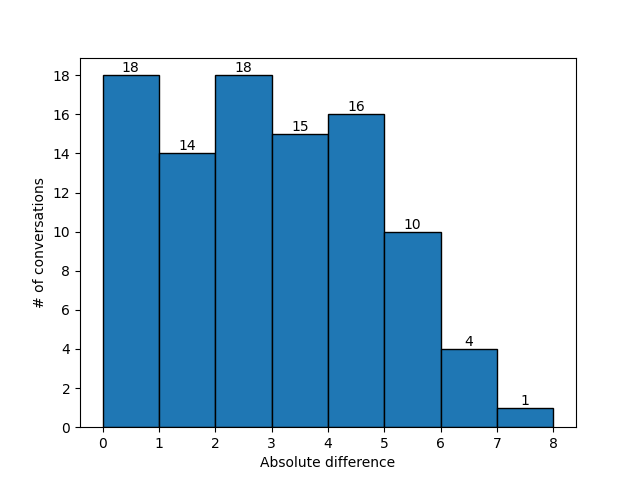}
  \caption{Absolute differences between the true scores and predictions for the 96 test conversations at K=20.}
  \label{fig:pred_diff}
  \vspace{-1.5em}
\end{figure}


Table \ref{tab:res} presents the regression results for different values of $K$ when contrastive or cross-entropy loss is employed. It is observed that the conversation embeddings learned with cross-entropy loss fail to predict TES scores. However, using the contrastive loss can encourage the conversation encoder to learn the embeddings that can predict TES scores effectively for all values of $K$ except $K=10$. The training data with $K=10$ may not be sufficient to train the network well. The best regression performance is achieved at $K=20$, where the mean absolute difference between the true scores and predictions is $2.93$\textpm$1.88$. Figure~\ref{fig:pred_diff} shows the distribution of the absolute differences for the $96$ test conversations. $81$ conversations have an absolute difference below $5.0$. Thus, the predicted results are considered reliable since empathy scores range from $9$ to $63$.

In addition, the conversations labeled by $[C_6,...,C_{25}]$ and $[C_{132},...,C_{151}]$ are selected to compose the training groups with smaller differences in empathy scores. The design of training data is denoted as $20(5)$ in Table \ref{tab:res}. Similarly, another design of training data denoted as $20(10)$ is also tried in the experiments. Both $20(5)$ and $20(10)$ have much worse $R^2$ than $K=20$. This suggests that the differences in empathy scores between the two training groups significantly impact the model performance.

From the above results, we can conclude that selecting counseling conversations with polarized ratings of empathy is effective for training the Siamese conversation encoder when using contrastive loss. Training with contrastive loss effectively encourages the conversation encoder to learn intra-group similarity and inter-group dissimilarity between the conversations from the low-empathy and high-empathy groups. Additionally, the paired nature of data in contrastive loss effectively increases the data samples during training. For instance, with $K=20$, only $40$ samples are available for cross-entropy loss training. In contrast, contrastive loss training allows access to $780$ $(2K^2-K)$ samples due to the pairing mechanism.





\section{Conclusion}
\label{sec:conc}


In this work, we propose to train the Siamese conversation encoder by using counseling conversations with polarized subjective ratings of therapist empathy. The use of contrastive loss is effective in encouraging the conversation encoder to learn discriminative embeddings for low-empathy and high-empathy conversations. Results show that the distances between the conversation embeddings are positively correlated with the differences in their empathy scores. In addition, the embeddings can be used to reliably predict the empathy level of therapists. The predicted empathy scores can serve as a reference for human raters, for example, to quickly identify counseling sessions with low scores for follow-up actions. We presume that the use of a conversation encoder can provide a solution to the problem of how to map a long conversation to a global score based only on acoustic features. Furthermore, the use of contrastive losses offers a solution to problems where only limited annotated data is available in similar tasks.

\section{Acknowledgement}
\label{sec:ack}

This research is partially supported by the Sustainable Research Fund of the Chinese University of Hong Kong (CUHK) and an ECS grant from the Hong Kong Research Grants Council (Ref.: 24604317).


\bibliographystyle{IEEEtran}
\bibliography{mybib}

\end{document}